\documentstyle[12pt]{article}

\def\intl{\int\limits}
\def\di{\displaystyle}
\def\&{&\di}
\def\bg{\begin{eqnarray}\begin{array}{rcl}\displaystyle}
\def\eg{\end{array} &\di    &\di   \end{eqnarray}}
\def\bm#1{\begin{eqnarray}\begin{array}{#1}\di} 
\def\bmo#1{\begin{eqnarray*}\begin{array}{#1}\di} 
\def\eg{\end{array} &\di    &\di   \end{eqnarray}}
\def\bgo{\begin{eqnarray*}\begin{array}{rcl}\displaystyle}
\def\ego{\end{array} &\di    &\di \nonumber  \end{eqnarray*}}

\def\btensor#1#2{\renew\left#1\begin{array}{#2}\di}
\def\etensor#1{\end{array}\right#1}

\def\ha{{1\over 2}}
\def\tp{\tilde{p}}

\def\d{{\mbox d}}

\def\Tr{\mbox{Tr}}

\def\id{1\!\mbox{l}}

\def\ov{\over}

\def\pa{\partial}
\def\al{\alpha}
\def\pr{\prime}
\def\dr{{D\!\llap{/}}}

\def\CD{{\cal D}}

\def\CF{{\cal F}}
\def\CG{{\cal G}}
\def\CH{{\cal H}}
\def\CL{{\cal L}}

\def\CP{{\cal P}}

\def\pan{\par\noindent}
\newcommand{\mtxt}[1]{\quad\hbox{{#1}}\quad}

\date{\today}

\def\rene{\renewcommand{\arraystretch}{1.8}}
\def\renew{\renewcommand{\arraystretch}{1}}
\rene

 \newcommand{\mysection}[1]{\section{#1}\setcounter{figure}{0}
                    \setcounter{table}{0}\setcounter{equation}{0}}

\begin{document}

\begin{titlepage}

\parindent=12pt
\baselineskip=20pt
\textwidth 15 truecm
\vsize=23 truecm
\hoffset=0.7 truecm

\begin{flushright}
   FSUJ-TPI-17/96 \\
      \end{flushright}
\par
\vskip .5 truecm
\large \centerline{\bf Polyakov-Loops and Fermionic Zero Modes in $QCD_2$
on the Torus} 
\par
\vskip 1 truecm
\normalsize
\begin{center}
{\bf U.~G.~Mitreuter}\footnote{e--mail:
  Mitreuter@tpi.uni-jena.de}, {\bf J.~M.~Pawlowski}\footnote{e--mail:
  Pawlowski@tpi.uni-jena.de}
{\bf and A.~Wipf}\footnote{e--mail:
  Wipf@tpi.uni-jena.de}\\
\it{Theor.--Phys. Institut, Universit\"at Jena\\ 
Fr\"obelstieg 1\\
D--07743 Jena\\
Germany}
\end{center}
 \par
\vskip 2 truecm
\normalsize
\begin{abstract} 
A direct derivation of the free energy 
and expectation values of Polyakov-loops 
in $QCD_2$ via path integral methods is given. 
The chosen gauge fixing
has no Gribov-copies and has a natural extension to four dimensions.
The Fadeev-Popov determinant and the 
integration over the space
component of the gauge field cancel exactly.
It only remains
an integration over the zero components of the gauge field 
in the Cartan sub-algebra. This way the Polyakov-loop 
operators become Vertex-operators in a simple
quantum mechanical model. 
The number of fermionic zero modes is related to
the winding-numbers of $A_0$ in this gauge. 
\end{abstract}
       
\vfill
       
\end{titlepage}

\mysection{Introduction}
A long standing and yet unsolved problem is 
proving quark confinement in QCD. 
An important first step in this direction would be to show confinement of
static quarks. In this way the problem reduces to understanding the
behaviour of electric flux strings in
pure $SU(N)$ gauge theories (without dynamical quarks). The relevant
observables are 
products of Wilson-loop operators \cite{wilson}. At finite temperature
the operators related to Polyakov-loops can be used to discriminate
between the confining and deconfining phases \cite{poloops}.

\noindent A rigorous construction and investigation of gauge theories in (3+1)
dimensions is beyond present days knowledge. (1+1) dimensional models
are much simpler and can be used as a testing ground to get more
insight into gauge theories on a sound mathematical basis.  In particular,
the unique and ambiguity-free gauge fixing described below can be
extend to $QCD_4$ \cite{tocome}.

\noindent Pure Yang-Mills theories in (1+1) dimensions are prototypes of
(almost) topological
field theories without propagating degrees of freedom. 
Nevertheless they have interesting
features, particularly in the large $N$ limit or/and on
multiple connected space-times \cite{witten}. 
The partition functions depend on $g^2V$, where $g$ is the
coupling constant and $V$ the volume of space-time, as well as 
invariants of the gauge group and topological
invariants of space-time. Polyakov-loops can be computed in both the strong and weak
coupling phase and the two phases are related by duality. It has been shown,
that the strong coupling expansion can be rewritten as a lower
dimensional string theory \cite{2}. When defined on Riemann
surfaces with non-zero genus they have degrees of freedom
related to the gauge group holonomy on the homology cycles of the
surface.  On cylindrical space-time they can be solved explicitly and
posess quantum mechanical degrees of freedom corresponding to the
eigenvalues of the Wilson loop operator which is winding  about the compact
space direction \cite{Langmann}. Such models
are also connected with one dimensional integrable quantum systems
\cite{polychronakos}.

\noindent The free energy $e^{-\beta F}=\Tr e^{-\beta H}$ at finite temperature 
$T=1/\beta$ is given by a path integral over
gauge fields on some manifolds $S^1\times M$ with
Euclidean time $x^0$ identified with $x^0+\beta$. (For discussion
of the path integral formulation of finite temperature gauge theory
see \cite{pisarski}.) Relevant gauge invariant order parameters 
are Polyakov-loop operators
\begin{equation}
\label{polloops}
P(x^1)=\Tr \,{\Gamma(\cal P}\,(\beta,x^1)),\mtxt{where}
{\cal P}(x^0,x^1)= \CP\,\exp \Big( i\intl_0^{x^0} A_0(\tau,x^1)\d \tau\Big).
\end{equation}
Here $\Gamma$ is the representation of the gauge group which acts on the
fermionic fields. For example, the two-point function
\begin{equation}
\label{freeenergy}
e^{-\beta F(x^1,y^1)}= \langle P(x^1) P^\dagger(y^1)\rangle_{\beta}
\end{equation}
yields the free energy $F(x^1,y^1)$ in the presence of a heavy
quark (in the fundamental representation) at $x^1$
and a heavy antiquark at $y^1$.  In the confining phase 
$F(x^1,y^1)$ increases for large separations of the quark-antiquark
pair and thus
$\langle P(x^1)P^\dagger (y^1)\rangle\to 0$.
In the deconfining phase the free energy 
reaches a constant value for large separations and thus
$\langle P(x^1)P^\dagger (y^1)\rangle\to const\neq 0$.
Inferring clustering we see that
$\langle P\rangle_\beta$ vanishes in the confining
phase but not in the deconfining one. In other words,
it is an order parameter for confinement.              
\noindent 
If we include massless dynamical fermions, the generating functional gains as
a factor the determinant of the Dirac operator. As a consequence
gauge field configurations which support fermionic zero modes do not
contribute to the partition function or to expectation values of
Polyakov-loops. Therefore, the question of
the number of zero modes for a given gauge field configuration is an 
important first step from pure to full $QCD$. 
               
\noindent In this paper we examine Yang-Mills theories on two dimensional tori.
They correspond to finite temperature
gluodynamics on a spatial circle. As shown by Grignani, Semenoff
and Sodano in an interesting paper \cite{1}\footnote{after we
  completed this work, Grignani et. al revised their paper, see
  \cite{grig2}},
  correlators of Polyakov-loop operators can be computed as correlators
in particular one dimensional models. For the case of pure gauge
theories (and Polyakov-loop operators in an arbitrary representation)
 one can explicitly solve these
quantum mechanical models. In contrast to other approaches we directly calculate, after an
appropriate gauge fixing, the partition function and the correlation
function $\langle P(x^1)P^\dagger (y^1)\rangle$ for arbitrary semi-simple gauge
groups. This will be a starting point for further
investigations concerning $QCD_4$ \cite{tocome}. 

\noindent In this paper we quantise Lie-algebra valued gauge fields
(non-compact $QCD_2$),whereas in \cite{witten,blau} the group
valued fields are quantised (compact $QCD_2$). 
As Hetrick has shown \cite{hetrick}, the non-compact theory has
additional spectral values connected with states, which lie on the
boundary of a Weyl chamber.
(Since these states lie in more than one chamber, they must be
added with an appropriate weight.) 
In the compact version  these states are projected to
zero-dimensional characters and are missing in the 
spectrum.  If these states
 are added to the partition function calculated
in \cite{witten,blau}, the results for the partition function $Z$ and expectation
values of products of Polyakov loop operators agree.
Due to this difference between compact and noncompact $QCD_2$, 
the numerical value of the string
tension for the static quark potential is different, but the physics
is qualitatively the same. 

\noindent In addition we
go beyond these results in that our approach leads to a simple relation between
the winding numbers and the number of fermionic zero modes.
           
\noindent In the first section we discuss the gauge fixing 
and topological questions connected with the definition of gauge
theories on $T^2$ (the corresponding results 
in $4$ dimensions are briefly sketched). In
the following two
sections we calculate the partition function and the
free energy of a static quark-antiquark pair. 
In the last section we derive a formula relating the number 
of fermionic zero modes to the winding numbers of
the gauge-fixed $A_{\mu}$. In the discussion we compare our results with those
of \cite{witten} and \cite{blau}. 
The appendices contain our Lie-algebra conventions
and a proof concerning antiholomorphic 
transition functions on the torus. 

\mysection{Gauge Fixing}
We view the torus $T^d$ as $R^d$ modulo a $d$-dimensional 
lattice, whose
points are denoted by $a,b,\dots$ with
coordinates $a_\mu=n_\mu L_\mu,\,n_\mu\in Z$ (no sum).
Matter fields and gauge potentials
on $R^d$ can be put on the torus
if they are (anti)periodic up to gauge 
transformations \cite{pierreHooft}
\bg
\psi(x+a) \& = \& (-1)^{n_0}\Gamma(U^{-1}_a(x))\psi(x),\\\di 
A(x+a) \& = \& U^{-1}_a(x)A(x)U_a(x)+iU^{-1}_a(x)\mbox{d}U_a(x),
\label{represent}
\eg
where the factor $(-1)^{n_0}$ enforces the finite
temperature boundary conditions for fermions $(L_0=\beta=1/T)$.
Since $\psi((x+a)+b)=\psi((x+b)+a)$,
the transition functions $U_a$ must obey 
the cocycle conditions \cite{pierreHooft}
\[
U_a(x)U_b(x+a)=U_b(x)U_a(x+b)Z_{ab},\qquad Z_{ab}=Z^{-1}_{ba},
\]
where the twists $Z_{ab}$ are 
in the kernel of $\Gamma$, i.e. $\Gamma(Z_{ab})=\id$. 
This kernel is
a subgroup of the center $Z$ of $G$.
For fermions in the fundamental representation no twists are allowed whereas
for fermions in the adjoined representation the twists can be any element
 of the center of $G$. 

\noindent Performing a (not necessarily periodic) gauge
transformation with $V(x)$, the new transition functions for 
the transformed fields are
\begin{equation}
\label{changetriv} 
\tilde{U}_a(x)=V^{-1}(x)U_a(x)V(x+a)
\end{equation}
The $\tilde{U_a}$ 
fulfill the cocycle condition with the same $Z_{ab}$ as the
$U_a$. Thus the twists are gauge invariant.
Note that the Polyakov-loop operators 
(\ref{polloops}) transform as
\bg
P(\vec{x})\longrightarrow \tilde P(\vec{x})=\Tr\big\{
V(0,\vec{x})V^{-1}(\beta,\vec{x})\,{\cal P}(\beta,\vec{x})
\big\},\mtxt{where} x=(x^0,\vec{x})
\eg
and are only invariant if $V(x)$ is periodic in time.
A twisted $G$-bundle over $T^d$ is uniquely 
characterized by the transition functions
modulo gauge transformation (\ref{changetriv}).

\noindent In the following we shall consider $2$-dimensional
gauge theories.
In $2$ dimensions and for 
simply connected gauge groups $G$ 
the $G$-bundles over $T^2$ are trivial
\footnote{This is not true, for
example, for $U(1)$ or $SO(3)$-bundles
over $T^2$ \cite{sachswipf}.}
(all Chern classes
are zero \cite{6}). 
Thus, in the untwisted case ($Z_{ab}=\id$) the 
transition functions can be chosen to 
be the identity. With twists
this is not true, but writing a
twist as $Z_{01}=Z=\exp(-2\pi i T)$,
we can always 
choose the transition functions as
\bg
\label{oldtrans}
U_{\beta}=\id\mtxt{and}U_L=e^{-2\pi i T x^0/\beta},
\eg
where $U_\beta$ relates the fields at $x^0$ and
$x^0+\beta$ and $U_L$ those at $x^1$ and $x^1+L$.

\noindent In explicit calculations
we must fix the
gauge. The field-dependent gauge transformation
which transform an $A_\mu$ into the gauge fixed form
may be non-periodic and thus lead to nontrivial 
transition functions.

\noindent After these general remarks we now discuss
the explicit gauge fixing.
Since Polyakov-loops only depend on $A_0$,
we shall choose a gauge for which $A_0$ 
is as simple as possible. Actually, with our
fixing $A_0$ will decouple
in the path integral and the 
expectation values of products
of Polyakov-loops can easily be 
calculated\footnote{similar fixings have been studied
in \cite{fixings}. After we discovered the gauge
fixing used in this work J. Fuchs pointed out to us that
E. Langmann et.al. found a very similar fixing.}. Below we prove that there is a 
(non-periodic) gauge
transformation which transforms any
$A_\mu$ into
\bg
A_0=A_0^c=2\pi H{x^1\over V}+A_0^{per}(x^1)\mtxt{,}
H\in \CL_\Gamma\mtxt{and}
\int\limits_0^{\beta}A_1^c (x^0,x^1)dx^0=C,
\label{gauge}
\eg
where $V=\beta L$ is the volume and 
we have introduced the following discrete lattice
in the Cartan-subalgebra $\CH$:
\bg
\label{fundamental}
\CL_\Gamma\equiv\Big\{H\in\CH\vert \Gamma\big(\exp(2\pi iH)\big)=
\id\Big\}.
\eg
In particular, for the fundamental and adjoint representations
$\Gamma=f$ and $\Gamma=adj$ we have
\bg
\CL_f=\Big\{H\in\CH\vert \exp(2\pi iH)=
\id\Big\}\mtxt{,}
\CL_{adj}=\Big\{H\in\CH\vert \exp(2\pi iH)\in Z\Big\}.
\label{twist1}
\eg
In (\ref{gauge}) $A^c$ is that part of $A$ which lies
in the Cartan subalgebra, $A_0^{per}$ is periodic
in $x^1$ and $A_1$ periodic in $x^0$. 
Below the constant  $C$ 
and $\int dx^1 A_0^{per}$ are further restricted
such that the gauge fixing (\ref{gauge}) becomes unique.

\noindent Note that the gauged fixed fields are not
periodic in $x^1$. Indeed, the transition
functions for the gauged fixed configurations are
\bg 
\label{newtrans}
\tilde U_{\beta}(x) =\id\quad\mtxt{and}\quad
\tilde U_L(x^0)=\exp\left\{-2\pi i H\frac{x^0}{\beta}\right\} 
\mtxt{with}H\in\CL_\Gamma .
\eg
Hence, the periodicity property of $A_1$ is given by
\begin{equation}
\label{gl26}
A_1(x^0,x^1+L)=e^{2\pi i \frac{x^0}{\beta}H} A_1(x^0,x^1) e^{-2\pi i
 \frac{x^0}{\beta}H}.
\end{equation}
Only $A_1^c\in\CH$ is periodic in $x^1$.

\noindent To prove, that (\ref{gauge}) can be achieved
we perform a gauge transformation
with 
\bg
\label{eich}
V(x^0,x^1) ={\cal
  P}(x^0,x^1){\cal P}^{-x^0/\beta}(\beta,x^1)W(x^1),
\eg  
where  ${\cal P}(x^0,x^1)$
has been defined in (\ref{polloops})
and $W$ diagonalizes ${\cal P}(\beta,x^1)$, i.e.
\bg
\label{diag}
{\cal P}(\beta,x^1)=W(x^1)\exp\{2\pi i H(x^1)\}\;W^{-1}(x^1).
\eg
This representation allows one to take
powers of ${\cal P}(\beta,x^1)$ and (\ref{eich}) 
becomes
\bg
\label{eich1}
V(x^0,x^1) ={\cal
  P}(x^0,x^1)\,W(x^1)\,\exp\Big(-2\pi i {x^0\ov \beta}H(x^1)\Big).
\eg  
Now it is easy to see that the gauge transformed 
$A_0$ reads
\bg
\label{atilde}
\tilde A_0={2\pi\ov \beta}H(x^1)
\eg
and hence depends only on $x^1$ and lies in the Cartan 
subalgebra.

\noindent By construction the gauge transformations (\ref{eich})
are periodic in time so that the Polyakov-loops are
unchanged, as required, and the transition
function in the time direction, $\tilde U_\beta$,
remains the identity, see (\ref{newtrans}).
To find $\tilde U_L$ we use
\bg
\label{shift1}
{\cal P}(x^0,x^1+L)=\exp(2\pi i T{x^0\ov \beta})
{\cal P}(x^0,x^1)
\eg
from which follows, that
\bg
\label{shift2}
\exp\{2\pi i H(x^1+L)\}&=&\exp\{2\pi i(T+H(x^1)\}\mtxt{and}
W(x^1+L)=W(x^1)
\eg
or equivalently that
\bg
H(x^1+L)=H(x^1)+H\mtxt{with}H\in\CL_\Gamma.
\eg
With (\ref{atilde}) and the consistency condition we end
up with the form (\ref{gauge}) for $A_0$.
The new transition function $\tilde U_L$ is easily
calculated from (\ref{changetriv}), with $U_L$
from (\ref{oldtrans}), $V$ from (\ref{eich1}) and
$V(x^1+L)$ from (\ref{shift1},\ref{shift2}). 
The result is the transition function
$\tilde U_L$ given in (\ref{newtrans}).

\noindent We have not yet fixed the gauge freedom completely.
Indeed, the residual gauge transformations are
\bg
\label{residual}
V(x)=w\cdot\exp\Big\{2\pi i (H_{per}(x^1)+H_0 {x^0\ov \beta}
+H_1{x^1\ov L})\Big\},
\eg
where all $H$'s are in the Cartan subalgebra and in addition
$H_{per}$ is periodic in $x^1$, $H_i\in \CL_f$
and $w$ is an element of normalizer$(\CH)/$centralizer$(\CH)
\cong\,$Weyl-group \cite{4}.
More explicitly, $w_\alpha$ acts on a
generator $H_\beta$ in $\CH$ as
\bg
w^{-1}_\alpha H_\beta w_\alpha=H_{\sigma_{\alpha}\beta},
\eg
where $\sigma_\alpha$ is the Weyl reflection related
to the root $\alpha$.
The $H_{per}$ part in (\ref{residual}) is fixed by imposing the 
second condition in (\ref{gauge}). The $H_i$-parts are fixed
if we further impose 
\bg
\label{fix2}
{\beta\ov L}\int\limits_0^L A_0^{per} dx^1\equiv{\beta\ov L}
\tilde C\in 2\pi\CH/\CL_\Gamma\quad\mtxt{and}
\quad{L\ov \beta}\int\limits_0^\beta A_1^c dx^0\equiv{L\ov \beta}C\in 
2\pi\CH/\CL_\Gamma.
\eg
It remains to fix the Weyl-transformations $w$
in (\ref{residual}). This can be done by imposing
the condition, that $\tilde C$ is in the first
Weyl-chamber.
However, the Weyl group is a finite group  
and permutes transitively and freely the Weyl
chambers, so the integration  over the $\tilde C$,
subject to (\ref{fix2}), is 
a multiple of the integration 
over the first Weyl chamber. If we consider normalised observables,
this overcounting cancels with that in the normalisation.

\noindent For later purposes it is important to note
that the transition functions for the gauge fixed
configurations possess abelian winding numbers. This can be seen 
as follows:
Since $\Gamma(\tilde U_L(x^0))=\Gamma(\tilde U_L(x^0+\beta))$
the map
\bg
\label{winding}
x^0\longrightarrow \Gamma(U_L(x^0)),
\eg
is a map from $S^1$ to $S^1\times\dots \times S^1$ 
($r$=rank$(G)$ factors)
and thus allows for $r$ integer winding  numbers.

\mysection{The functional integral}
In the following we decompose the Lie algebra valued gauge
potential (for conventions see the appendix) as follows
\begin{eqnarray}
A_0(x)\& = \&\sum_{\al\in\Delta} p^\al(x)H_\al
+\sum_{\varphi\in \Phi^+ }
a^{\varphi}(x)E_{\varphi}+\sum_{\varphi\in \Phi^+}
\bar a^{\varphi}(x)E_{-\varphi}\nonumber\\\di 
A_1(x)\& = \&\sum_{\al\in\Delta} q^\al(x)H_\al+\sum_{\varphi\in
\Phi^+}
b^{\varphi}(x)E_{\varphi}+\sum_{\varphi\in\Phi^+ }
\bar b^{\varphi}(x)E_{-\varphi},\nonumber
\end{eqnarray}
where $\Delta, \Phi^+$ and $\Phi^-$ denote the simple,
positive and negative roots, respectively
and $H_\al$ is the generator in the Cartan subalgebra $\CH$
belonging to the simple root $\alpha$. 

\noindent The gauge fixing conditions (\ref{gauge}) read
\bg
\Tr\,E_\varphi A_0=\Tr\,E_{-\varphi}A_0=
\Tr\,H_\alpha\big(\pa_0 A_0-{1\ov \beta}\int dx^0\pa_1A_1\big)=0
\label{ourgauge}.
\eg
For the gauge fixed configurations
(see (\ref{gauge})) we find for the
field strength
\bg
F_{01}=\sum_\al(\dot q- p^\pr)^\al H_\al
-i\sum_\varphi \big[M_\varphi b^\varphi E_\varphi
-(M_\varphi b^\varphi)^*E_{-\varphi}\big],
\qquad
M_{\varphi}=\big(i\pa_0+\sum_\al K_{\varphi\al}p^\al\big)
\eg
is hermitean.
Correspondingly the gauge fixed action reads
\bg
S =\frac{1}{8 g^2}\int \d x^0 \d x^1 \;\Tr\;F_{\mu\nu} F^{\mu\nu}
=\frac{1}{4g^2}\int \Big\{(\dot q-p^\pr,C\,
(\dot q -p^\pr))+\sum_\varphi 
{4\ov \varphi^2}(b^\varphi,M^2_\varphi b^\varphi)\Big\},
\eg
where $C=(C_{\al\beta})$ is the symmetric
Coxeter matrix and the $r$-component 
real vector fields $p$ and $q$ have entries $p^\al$
and $q^\al$, respectively.
The last scalar product containing the operators
$M_\varphi$ is a complex one, $(b,c)=\bar b\cdot c$.
To calculate the Fadeev-Popov determinant we observe that
the gauge variation of the gauge fixings $\Tr(E_{\pm \varphi}A_0)=0$
are
\bg
\delta_\theta \Tr E_{-\varphi}A_0={2i\ov \varphi^2}M_\varphi
\theta^\varphi\mtxt{and}
\delta_\theta \Tr E_{\varphi}A_0=-{2i\ov \varphi^2}(M_\varphi
\theta^\varphi)^*\mtxt{(no sum).}
\eg
Vanishing variations imply vanishing $\theta^\varphi$ and
then the variation of the remaining gauge fixings
simplifies to
\bg
\delta_\theta \Tr H_\al\Big(\pa_0 A_0+{1\ov V}\int dx^0\pa_1 A_1\Big)
=-C_{\al\beta}\Big(\pa_0^2+{1\ov \beta}\int dx^0\pa_1^2\big)\theta^\beta
\eg
Now we see, that for simply laced 
groups (for which the length of
all roots can be taken to be $2$)
the field-dependent Fadeev-Popov determinant 
coming
from the $\theta^\varphi$ cancel 
exactly against the functional integral over the
non-Cartan fields $b^\varphi$. Thus we obtain the following
partition function
\bg
Z&=&N\int\CD q (x) \CD p(x^1)\,\delta(\CF(A))
\det(C)\det(-\partial_0^2-
\frac{1}{L}\int\mbox{d}x^0\;\partial_1^2)^{r}\\
&\cdot&
\exp\Big\{{1\ov 4g^2}
\int(\dot q-p^\pr,C
(\dot q-p^\pr))\Big\}
\eg
with a normalization factor $N$.
Here $\delta(\CF(A))$ indicates the implementation of
the zero mode fixings (\ref{fix2}).
Since $A^c_1$ is periodic in $x^0$
and $A_0$ depends only on $x^1$,
the integration over $q^\al$ decouples completely
and we end up with
\begin{equation}
Z=N' \int \CD p(x^1)\exp
\left\{-\frac{\beta}{4g^2}\int (p^\pr,
C\,p^\pr)\,\mbox{d}x^1\right\}.
\end{equation}
For simplicity we restrict ourselves to $\Gamma=f$ (fermions in the
fundamental representation), $G=SU(N)$ 
with rank $r=N-1$ and choose the basis
\bg
E_i=E_{\alpha_i}=E_{i,i+1}\mtxt{and}H_i=[E_{\alpha_i},E_{-\alpha_i}]\mtxt{for\,
simple}\alpha_i,
\eg
where $E_{i,j}$ is the $N\times N$-matrix whose only 
non-zero entry is a $1$ in the $i$'th row and $j$'th
column. The step-operators belonging to the non-simple
roots are obtained by commutation of the $E_i$.
The center consists of the $N$'th roots of unity 
and is generated by the
\bg
\label{twist}
T_{\tau}={\tau\ov N}\;\hbox{diag}(1,1,\dots,1-N),\qquad
\tau=0,1,\dots N-1.
\eg
Then $C=K$ has $2$'s on the diagonal and $-1$ on the two
off-diagonals above and below the diagonal.
We can decompose the $p=(p^1,\dots,p^r),\;\,p^i=p^{\al_i}$ as follows
\bg
\label{decompos}
p(x)={1\ov\beta}\big[\,\tp(x)+h\big]+\frac{2\pi n}{V} x\qquad
(x=x^1),
\eg
where we have separated the constant part $h=(h^1,\dots,h^r)$ 
of the periodic 
piece, so that the 
$\tp^i$ are periodic in $x$ and integrate to zero. 
Since $\Gamma(\exp[2\pi \vec n\cdot H])=\id$,
the $\vec n$ lie in $Z^r$ for matter in the fundamental
representation and $n^i\in \tau/N+Z$ for matter
in the adjoint representation. In the explicit
calculations below we assume that there are no twists.
We shall give the corresponding results for
the twisted case at the end of the next section.

Inserting the decomposition (\ref{decompos})
we find for the partition function
\bg
\label{sum}
Z\sim \int\CD\tp \,d^r h\,
\exp \Big\{\frac{1}{4g^2\beta}\int (\tp,K\partial_1^2
\tp)dx\Big\}\cdot 
\sum_{\vec n\in Z^r}\exp\Big\{-\frac{\pi^2}{g^2V}
(\vec n,K \vec n)\Big\}.
\eg
Due to the fixing of the time dependent residual gauge freedom
(\ref{residual}) the $h^i$-integrations
are restricted to the interval  $[-\pi,\pi]$. 
        
\noindent Using zeta-function regularisation
the Gaussian integration over $\tp^i$ yields
\[
\left({\det}^\pr\left[\frac{-K\partial^2}{2g^2\beta}\right]
\right)^{-1/2}=
{\det}^{1/2}{K\ov 2 \beta L^2 g^2}.
\]
After a Poisson-resummation in (\ref{sum}) 
we end up with
\begin{equation}
Z\sim \sum_{\vec m\in Z^r}\exp \{-g^2 V (\vec m,K^{-1}\vec m)\}
\end{equation}
with the inverse of the Cartan matrix 
\begin{equation}
(K^{-1})_{ij}=\frac{1}{N}(N-j)i, \;\;\;\mbox{for}\quad i\leq j,
\quad(K^{-1})_{ij}=(K^{-1})_{ji}.
\end{equation}

\mysection{Calculation of Polyakov-loops}
For the gauge fixed configurations and $\Gamma=f$ the Polyakov-loops 
(\ref{polloops}) simplify to
\bg
P(x)=\Tr \exp\left\{i\beta\, p(x)\cdot H\right\}
=\sum_{k=1}^N \exp\left\{i\beta [p^k(x)-p^{k-1}(x)]\right\},
\eg
where $p^0\equiv p^N\equiv 0$. We get for the expectation value of
the product of two Polyakov-loops
\bg
\langle P(x)P^\dagger(y)\rangle\!=\!\frac{1}{Z}\int\CD \tp\,d^rh\,
\sum_{\vec m} \exp\left\{
-\frac{\pi^2}{g^2V}(\vec m,K \vec m)+
\frac{1}{4g^2\beta}\int (\tp,K\partial^2\tp)\right\}P(x)
P^\dagger(y).
\eg
After integration of the $h^i$ only the
diagonal elements in the double sum (coming
from the $2$ Polyakov-loop operators) contribute and
\bgo
\langle PP^\dagger \rangle \&\!\!\! = \&\!\!\! {1\ov Z}\sum_{k=1}^N \sum_{\vec m}\exp \left\{
-\frac{\pi^2}{g^2V}(\vec m,K\vec m)+2\pi i m^k\xi-
2\pi i m^{k-1}\xi\right\}\\\di 
\&\&  \int\CD\tp\exp\left\{ 
\frac{1}{4g^2\beta}\int (\tp,K\partial^2\tp)\mbox{d}x
+i[\tp^k(x)-\tp^k(y)]-i[\tp^{k-1}(x)-\tp^{k-1}(y)]\right\}, \nonumber
\ego
where we have introduced $\xi=(x-y)/L$ (recall that $x\equiv x^1$).
To calculate the functional integral over the $\tp^i$ we need the 
zero mode truncated Greens function 
of $-1/2g^2\beta\cdot K\pa^2$ which is 
\bg
G(x,y)=
K^{-1}\Delta(x,y),\mtxt{where}\Delta(x,y)=g^2V\Big(\xi^2-\vert
\xi\vert +{1\ov 6}\Big) \ \ \mbox{for}\ \  \xi\in[-1,1].
\label{green}
\eg
Now we perform a Poisson resummation of the $N-1$ sums 
and calculate the Gaussian functional integral over the 
periodic $\tp^i$. We emphasize that there are no
zero mode problems, as it must be for a complete gauge
fixing. The result is
\bg
\langle P(x)^\dagger P(y) \rangle\& \!\! = \& \!\! {1\ov Z}\sum_{k,\vec m}
\exp \left\{ -g^2 V
\left(m_i-\xi[\delta_{ik}-\delta_{i(k-1)}]\right)K^{-1}_{ij} 
\left(m_j-\xi[\delta_{jk}-\delta_{j(k-1)}]\right)\right\}\\\di 
\& \&  \exp\left\{\frac{N-1}{N}[\Delta(x,y)-\Delta(0,0)]\right\}
\eg
where $K^{-1}_{0i}=K^{-1}_{Ni}=0$. Thus the expectation value of the product of two Polyakov
loops reads 
\bm{c}
\langle P(x)P^\dagger (y) \rangle= {1\ov Z}\sum_{k,\vec m}\exp
\left\{-g^2V \left(\vec m,K^{-1}\vec m-2\xi m_i
\left[K^{-1}_{ik}-K^{-1}_{i(k-1)}\right]+\vert\xi\vert\frac{N-1}{N}\right)\right\}

\eg 
for $\xi\in [-1,1]$. For $SU(2)$ this simplifies to 
\begin{equation}
\label{343}
\langle P(x)P^\dagger (y) \rangle =\frac{2}{Z}\sum_{m\in Z}
\exp \left\{ -\frac{g^2V}{2}\left[
m^2-2\xi m+\vert\xi\vert\right]\right\}.
\end{equation}
The free energy for the static quark-antiquark
pair in the fundamental representation is
gotten from (\ref{freeenergy}). $Z$ is given by $\langle P(x)P^\dagger(x) 
\rangle=1$. For large
separations of the pair we find for the free energy for $SU(N)$
\begin{equation}
\label{gl47}
\lim_{L\to\infty} F(x,y)=g^2\frac{N-1}{N} |x-y|.
\end{equation}
We conclude this section with the analogous results for
the free energy of a static quark-antiquark
pair in the adjoint representation, for which
$adj[\exp(2\pi i H)]=\id$. In this case the
gauge fixed $A_0$ has the decomposition
\begin{equation}
A_0=\sum p^kH_k\mtxt{with}
p^k(x)={1\ov \beta}[\,\tp^k+h^k]+{2\pi\ov V}(k{\tau\ov N}+m^k)x
,\qquad \tau=0,\dots,N-1\end{equation}
and the Polyakov-loop is
\bmo{rclcrcl}
P(x) \& = \& \Tr\,\Gamma_{adj}\left(\exp\left(i\int_0^\beta
A_0(\tau,x)\d\tau\right)\right) \& = \& \Tr \exp\left(i\int_0^\beta
A_0^k(\tau,x)\Gamma^*_{adj}(H_k)\d\tau\right). 
\ego
where $\Gamma^*$ is the Lia algebra representation induced by $\Gamma$.
Now one proceeds as in the untwisted case. One obtains with 
$\tilde m=\sum_l m_l l$ and $\xi\in[-1,1]$ 
\bg
\langle P(x)P^\dagger(y)\rangle\& = \&\frac{1}{Z}\left[r^2+2N\sum_{p=1}^r\sum_{j=1}^p
\sum_{\vec m}
\exp \left\{ -g^2 V 
\left(\vec m K^{-1}\vec m-2\xi
\sum_{i=j}^p m_i\right)\right\}\right.\\\di 
\& \& \left.\exp\left\{-g^2V\vert\xi\vert 2(p-j+1)\right\}
\sum_{n=-\infty}^{\infty}\delta_{nN,\tilde m}\right].
\label{pol}\eg 
For $SU(2)$ (\ref{pol}) simplifies to
\bg
\langle P(x)P^\dagger(y)\rangle = 1+\frac{4}{Z}\sum_{m}\exp 
\left\{ -2g^2 V\left(m^2-2m\xi+|\xi|\right)\right\}.
\label{pol2}\eg
For large separations of the pair we get for the free energy in the
twisted case for $SU(N)$ 
\begin{equation}
\lim_{L\to\infty} F(x,y)\sim
-\frac{1}{\beta}\frac{2N}{r^2}\sum_{p=1}^r\sum_{j=1}^p
\exp\left\{-g^2\beta|x-y|2(p-j+1)\right\}.
\end{equation}
For $\frac{1}{g^2\beta}\ll |x-y| \ll L$ the free energy becomes zero and 
due to the cluster decomposition theorem the expectation value of one 
Polyakov-loop operator is one in agreement with \cite{blau}.

\mysection{Zero Modes of the Dirac-operator}
In this section we characterize and count
the number of zero modes (in the fundamental
representation) of $\dr$ for gauge theories on $T^2$. We will show 
that the number of fermionic zero modes for gauge fixed 
configurations $A_\mu$ with
transition functions (\ref{newtrans}) is just
\bg
n_0=\Tr\vert H\vert.
\label{zeronumber}
\eg
The analogous result on $S^2$ has been derived in \cite{mickel}. To
prove (\ref{zeronumber}) we introduce the 
complexification $G^c$ of $G$
and assign to each gauge fixed $A$ (in the gauge (\ref{gauge})) 
the set of $G^c$-valued prepotentials
\bg
\CG_A \& = \& \{g(z,\bar z)\in G^c|\ A_z = i g^{-1} \partial_z g\}
\label{definition1}
\eg
with $A_z:=\ha(A_0-i A_1)$ and $z=x^0+i x^1$.
Since the $G$-bundles over $T^2$ are trivial each
gauge fixed $A$ is a gauge transform
of a periodic potential $A^p$,
\bg
A_z=V_A^{-1}A_z^p V_A+iV_A^{-1}\partial_z V_A
\eg
and  the prepotentials belonging to $A$ are
\bg
\CG_A=\Big\{g(z,\bar z)=h(\bar z)\cdot g_A(z,\bar z)V_A\vert\;
g_A(z,\bar z)=\CP \exp \big\{i\int\limits_z^0 
\d u A^p_z(u,\bar z)\big\}\Big\}.
\label{explicit}\eg
>From the periodicity of $A^p$ and the known transition
functions (\ref{newtrans}) of $A$ one can read off the 
nonperiodicity of the $V$:
\bg 
V_A(x^0+n\beta,x^1+mL)=V_A(x^0,x^1)e^{-2\pi i mH x^0/\beta}
\eg
Now we classify the non-periodicity of the
prepotentials $g$ in (\ref{definition1},\ref{explicit}). 
Since $A^p$ is periodic it follows that
\bg 
g(z+n\beta+im L,\bar z +n\beta-imL) \& =\& h_{nm}(\bar
z)g(z,\bar z)e^{-2\pi i m \frac{x^0}{\beta}H}
\eg
The antiholomorphic $h_{nm}$ are 
transition functions of homomorphic vector bundles
over the 2-dimensional torus\footnote{
e.g. $g=g_AV_A$ has transition function
$
h_{nm}(\bar z)=\CP\exp\big\{-i\intl_0^1
A^p_z\big(-\tau (n\beta+imL),\bar z\big)\cdot
(n\beta+imL)\d\tau\big\}.$}
and must obey the cocycle conditions
\bg
h_{nm}(\bar z+p\beta-iqL)h_{pq}(\bar z)=
h_{pq}(\bar z+n\beta-imL)h_{nm}(\bar z).
\label{comcocycle}
\eg
            
\noindent To continue, we note that if
$g\in \CG_A$ has transition functions $h_{nm}(\bar z)$, then
$h(\bar z)g\in \CG_A$ has transition functions
\bg
\tilde h_{nm}(\bar z) \& =\& h(\bar z+n\beta-imL) h_{nm}(\bar z)
h^{-1}(\bar z).
\label{transformation}\eg
Using this gauge freedom we can always find
a representative in $\CG_A$ such that
$h_{n0}(\bar z)=\id$. 
To see that we write the $h_{n0}$ as
\bg
h_{n0}=\CP\exp\Big\{i\intl_{\bar z}^{\bar z+n\beta}\d\bar u\
a(\bar u)\Big\}.
\eg
Then
\bg
h(\bar z)=\CP\exp\Big\{i\int\limits_{\bar z}^0\d\bar u\
a(\bar u)\Big\}
\eg
transforms the $h_{n0}$ into the identity, as required.

\noindent It follows from the cocycle conditions (\ref{comcocycle}) 
that the remaining nontrivial transition functions must
be periodic in time, 
\bg
h_{0m}(\bar z +n\beta)=h_{0m}(\bar z).
\label{prove1}
\eg
In the appendix B we shall prove, that the $h_{0m}$ 
can be written as
\bg
h_{0m}(\bar z)= 
V_L^{m^2}\cdot e^{2\pi i m\frac{\bar z}{\beta}H_A}\cdot  
\CP\exp\Big\{ i\intl_{
\bar z}^{\bar z-imL}\d\bar u\ b_p(\bar u)\Big\},
\label{prove2}
\eg
where $b_p$ is periodic in $x^0$, $H_A$ lies in the Cartan 
subalgebra and is
quantized, $\exp(2\pi i H_A)=\id$, and
\bm{rclcrcl}
V_L \& = \& e^{\pi\frac{L}{\beta}H_A} \& \mbox{and} \& [H_A,b_p(\bar u)]
\& = \& 0.
\eg
Now we make a further gauge transformation with
\bg 
h(\bar z)=\CP\exp\big\{i\intl_{\bar z}^0\d \bar u
\ b_p(\bar u)\big\}.
\eg
The new transition functions $h_{nm}$ read
\bm{rclcrcl}
h_{nm}(\bar z)\& =\& V_\beta^n\cdot V_L^{m^2}\cdot 
e^{2\pi im \frac{\bar z}{\beta}H_A}
\& \mtxt{with} \& V_\beta \& = \& \CP\exp\Big\{i\intl_\beta^0b_p(\bar u)
d\bar u\Big\}.
\eg
Setting $V_\beta = 
e^{v_\beta}$ we can factorize $g$ as
\bgo
g(z,\bar z)=e^{v_\beta x^0/\beta}e^{\pi H_A(x^1)^2/V}\tilde g(z,\bar z).
\ego
The non-periodicity of $\tilde g$ is simply
\bgo
\tilde g(z+n\beta+imL,\bar z+n\beta-imL) \& = \& e^{2\pi i m
  \frac{x^0}{\beta}H_A}
\cdot \tilde g\cdot e^{-2\pi i m\frac{x^0}{\beta}H}.
\ego 
In terms of $\tilde g$ the gauge fixed potential reads
\bg
A_z =\tilde g^{-1} \Big(i\partial_z+A_I\Big) \tilde g,\quad\quad\mtxt{where}
A_I={\pi x^1\ov V}H_A+{i\ov 2\beta}v_\beta
\eg 
is an abelian instanton potential, $[A_I,v_\beta]=0$.\pan
Now it is easy to see that the Dirac-operator $\dr\,(A)$ can
be related to the one in the instanton background as
\bm{ccc}
\label{dirac1}
\dr\,(A) = \tilde G^\dagger \dr\,(A_I) \tilde G, \& 
\tilde G=\btensor{(}{cc}\di \tilde g^{\dagger -1} \& 0 \\\di 0 \& \tilde g
\etensor{)}, \& \dr\,(A_I)= 2\btensor{(}{cc} 0 \& \partial_z-iA_I
\\\di  \partial_{\bar z}-iA_I^\dagger\& 0 \etensor{)}.
\eg
It follows at once that
\bg
\psi_0 = \tilde G^{-1}\tilde\psi_0
\label{zero1}
\eg
is a zero mode of $\dr\,(A)$ if $\tilde \psi_0$ is a zero mode
of $\dr\,(A_I)$. \pan
Let us calculate the left-handed ($\gamma_5=-1$) zero modes in
the instanton background $A_I$. Comparing $A_I$ with the
general gauge fixed form (\ref{gauge}) we see that $H_A\equiv H$.
The Dirac-eqation reads
\bgo
\big(\pa_z-i{\pi x^1\ov V}H+{1\ov 2\beta}v_\beta\big)\tilde\psi_0=0
\ego
and is solved by the spinor fields
\bg
\tilde \psi_0(x^0,x^1)=e^{-\pi (x^1)^2H/V-iv_\beta x^1/\beta}\chi(\bar z).
\label{zero2}
\eg
The zero modes must fulfill the boundary conditions (\ref{represent})
with transition function (\ref{newtrans}) so that
$\chi$ must be antiperiodic in time and 
\bg
\chi(\bar z-iL)=e^{2\pi iH\bar z/\beta +L(\pi H-iv_\beta)/\beta}
\chi(\bar z).
\label{hhh}
\eg
Thus $\chi$ can be expanded as
\bg
\chi(\bar z)=\sum_n e^{\pi i(2n+1)\bar z/\beta}a_n,
\label{zero3}
\eg
where the Fourier-coefficients $a_n$ transform according
to the fundamental representation of $G$.
To proceed we use the fact that $H$ commutes with the
Dirac-operator and can be diagonalized, $Ha_n=ma_n$.
Since $v_\beta$ commutes with $H$ it leaves the
subspace on which $H=m$ invariant. On this subspace
(\ref{hhh}) translates into
\bg
a_n=e^{\pi(m-1-2n)L/\beta}e^{-iv_\beta L/\beta}a_{n-m}.
\label{zero4}
\eg
Now we see that we can choose the vectors $a_1,\dots,a_{m}$
freely, so that there are $m\cdot$(degeneracy of $m$)
normalizable zero modes if $m$ is positive. Repeating
the same procedure for the right-handed zero-modes, for
which $m$ must be negative, we end up with the following
formula for the number of zero-modes
\bg
n_0=\Tr\vert H\vert.
\eg
The explicit zero modes are given by 
(\ref{zero1},\ref{zero2},\ref{zero3}) where the
$a_n$ are determined by the recursion relations
(\ref{zero4}). This way one finds, that the
zero modes in the instanton backgrounds
are theta-functions ( similar to the abelian
Schwinger model \cite{sachswipf}).

\mysection{Discussion}
In this paper we have given a simple 
derivation for the expectation value of
Polyakov-loops in $QCD_2$ at finite temperature.
For the simplest case, $G=SU(2)$ and matter in
the fundamental representation, i.e. with untwisted
gauge fields, the interaction energy between
two widely separated external sources is 
\bm{rcl}
F(x,y) \& \stackrel{L\rightarrow \infty}{\rightarrow}\& 
{g^2\ov 2}\vert x-y\vert.
\eg
If we twist the gauge fields and thus introduce magnetic
flux quanta we get 
\bm{rcl}
F(x,y) \& \stackrel{L\rightarrow \infty}{\rightarrow}\&
-\frac{4}{\beta}\exp\{-2g^2\beta|x-y|\}
\eg
A similar behaviour is found for the higher groups and maximally twisted
and untwisted fields. In the untwisted case 
we get a confining potential, whereas in the twisted case
the potential $F(x-y)$ decays exponentially to a
constant, which is to be interpreted as screening of the external
charges. In figure~1 we show $F(x,y)$ for abitrary
$|x-y|/L\in[0,1]$ for $SU(2)$ in the untwisted case. 
In figure~2 we show $F(x,y)$ for $SU(2)$ in
the twisted case. For large volume $F(x,y)/V$
tents to zero everywhere. 

\noindent The string tension (\ref{gl47}) in non-compact $QCD_2$ on the torus is
different from the one in compact $QCD_2$ \cite{witten,blau,grig2}.
For example, for the partition function (for $SU(2)$) the two
quantisations differs on the $n=0$ contribution in
\[
Z=\sum_n n^{2-2g} e^{-g^2Vn^2}
\]
In \cite{tompson} it has been argued, that the $n=0$ term is absent
for $g\neq 1$, but on the torus there is no way to decide, which of
the two quantisations is the correct one.
Since in our path integral quantisation we do not need to fix the Weyl
symmetry, we get twice the result of non-compact $QCD_2$.
Therefore the 'zero representation' of Hetrick \cite{hetrick}
must be added to the partition function of compact $QCD_2$ with
a factor 1/2. Another argument for a factor 1/2 is,
 that the corresponding state lies on the boundary
of the Weyl chamber and hence belongs to two chambers simultanously.
In order to avoid double counting, we need the factor 1/2. These weights
are also present in the calculation of expecation values of Polyakov
loops.

\noindent To check the cluster decomposition theorem one must compute
the expectation value of one Polyakov-loop operator for the twisted
and untwisted case. We get $\langle P\rangle_f=0$ and $\langle
P\rangle_{adj}=1$. This agrees with calculations of expectation
values of homologically nontrivial Wilson-loops on genus one Riemann
surfaces done in \cite{blau}.

\noindent In a second part we derived an explicit formula relating 
the $r$ winding numbers
of the gauge fixed configurations to the total number
of zero modes. Indeed, the number of zero modes is just
the product of the winding numbers. This is a nontrivial result
and we believe it is new. It goes much beyond the well known index theorem,
which is trivial in two dimensions.
\pan
We would like to point out, that the gauge fixing
introduced in section 2 has a natural extension to
higher dimensions. For example in $4$ dimensions
the generalization of (\ref{gauge}) reads
\bm{rclrclrclrcl}
A_0\& =\& 2\pi H_0{x^1\over L_0L_1}+\tilde A^c_0,\& 
A_1\& =\& \tilde A_1, \& 
A_2\& =\& 2\pi H_2{x^3\over L_2L_3}+\tilde A_2,\& 
A_3\& =\& \tilde A_3
\eg
where the Cartan-pieces of the $\tilde A_\mu$ are
constrained by
\bg
\tilde A_0^c=C_0(x^1,x^2,x^3)&\mtxt{,}&\int dx^0\tilde A_1^c=C_1(x^2,x^3)\\
\int dx^0dx^1\tilde A_2^c=C_2(x^3)&\mtxt{,}&\int dx^0dx^1dx^3\tilde A_3^c=
C_3.
\eg
The constant parts of the $C_\mu\in \CH$ are further restricted
to avoid Gribov copies. Actually a slight modification of
this gauge fixings can be achieved 
in all instanton sectors for $G=SU(N>2)$ and in the sectors
with even instanton numbers for $SU(2)$ \cite{tocome}.
\subsection*{Acknowledgement}
We would like to thank Chris Ford, Christoph Adam and Nico Giulini for 
numerous discussions, J. Fuchs for drawing our attention to 
related work of E.~Langmann et. al \cite{fixings} and Matthias Blau
for discussions on the difference between compact and non-compact
$QCD_2$.

\begin{appendix}
\section*{Appendix}
\mysection{Conventions}
In sections 2,3 and 4 we used the Chevalley basis \cite{4}
\bg
[H_{\alpha},H_{\beta}]= 0,\quad\forall\,\alpha,\beta\in \Delta &\mtxt{,}&
[H_{\alpha},E_{\beta}]= K_{\beta\alpha}E_{\beta}\quad\forall\,\alpha,\beta\in
\Delta\\\di
[E_{\alpha},E_{-\alpha}]=H_{\alpha},\quad\forall\,\alpha\in\Delta&\mtxt{,}&
[E_{\alpha},E_{\beta}]=E_{\alpha+\beta},\quad\forall \,\alpha+\beta \in
\Phi^+ \\\di
[H_{\alpha},E_{\beta+\gamma}]=
\big(K_{\beta\alpha}+K_{\gamma\alpha}\big)E_{\beta+\gamma}
&,&\quad\forall\, \alpha,\beta,\gamma\in\Delta,\quad\beta+\gamma \in \Phi^+ 
\label{group1}
\eg
where $\Delta$ is the set of simple roots and $\Phi^+ $ the set of
positive roots. The Cartan matrix and the symmetric Coxeter matrix
are given by
\bg
K_{\alpha\beta}=\frac{2\langle\alpha,\beta\rangle}{\langle\beta,
\beta\rangle}\mtxt{and}
C_{\alpha\beta}=\frac{4\langle\alpha,\beta\rangle}{\langle
\beta,\beta\rangle\langle \al,\al\rangle}.
\label{group2}
\eg
In the body of this paper we used $K_{\alpha\varphi}$ for
simple $\al$ and positive $\varphi$. This 'extension' of
the Cartan matrix is defined as for simple roots, see (\ref{group2}).
For the traces we get
\bg
\Tr (H_{\alpha}H_{\beta})=\frac{2}{\alpha^2}K_{\alpha\beta}\mtxt{,}
\Tr (E_{\alpha}E_{\beta})=
\frac{2}{\alpha^2}\delta_{\alpha,-\beta}\mtxt{and}
\Tr (H_{\alpha}E_{\beta})=0,
\eg
where $\Tr$ is the usual matrix trace multiplied by an appropriate
normalisation constant which ensures $|\alpha_{long}|^2=2$. 

\mysection{Proof of (\ref{prove2})}\label{proof2}
To prove (\ref{prove2}) we rewrite $\tilde h_{0m}$ as a
path ordered exponential. We define $b(\bar z)$ by 
\bg 
\tilde h_{0m}(\bar z) \& = \& \CP\exp\Big\{i\intl_{\bar z}^{\bar z-imL}\d\bar
u\ b(\bar u)\Big\}.
\eg 
The periodicity of $\pa_{\bar z}h_{0m}$ in $x^0$ translates into
\bg
b(\bar z+p\beta-imL)-b(\bar z+q\beta-imL)=
h_{0m}(\bar z)\Big[b(\bar z+p\beta)-b(\bar z+q\beta)\Big]
h_{0m}^{-1}(\bar z)
\label{condition}
\eg
for arbitrary integers $p,q$ and $m$.
It follows at once, that $b$ must be periodic in $x^0$,
up to a linear term,
\bg
b(\bar u)=b_1 \bar u+b_p(\bar u).
\eg
From (\ref{condition}) it follows that
$h_{0m}$ lies in the little group of
$b_1$, 
\bg
h_{01}(\bar z)b_1 h_{01}^{-1}(\bar z)=b_1
\eg
As a consequence $b_1$ commutes with $b_p$ and
\bg 
h_{0m}=e^{-(m^2 L^2 b_1/2+imL b_1 \bar z)} \CP\exp\Big
\{i\intl_{\bar z}^{\bar z-imL}\d\bar u\ b_p(\bar u)\Big\}. 
\eg
Since the $h_{0m}$ are periodic in $x^0$ with period
$\beta$, the constant $b_1$ has to be $2\pi H_A/V$ with $\exp(2\pi i
H_A)=1$. Thus $H_A$ is diagonalizable and can take it to be in the Cartan
subalgebra.
\end{appendix}

\newpage
\unitlength=1cm
\begin{picture}(18,7)
\put(0,0){
\includegraphics{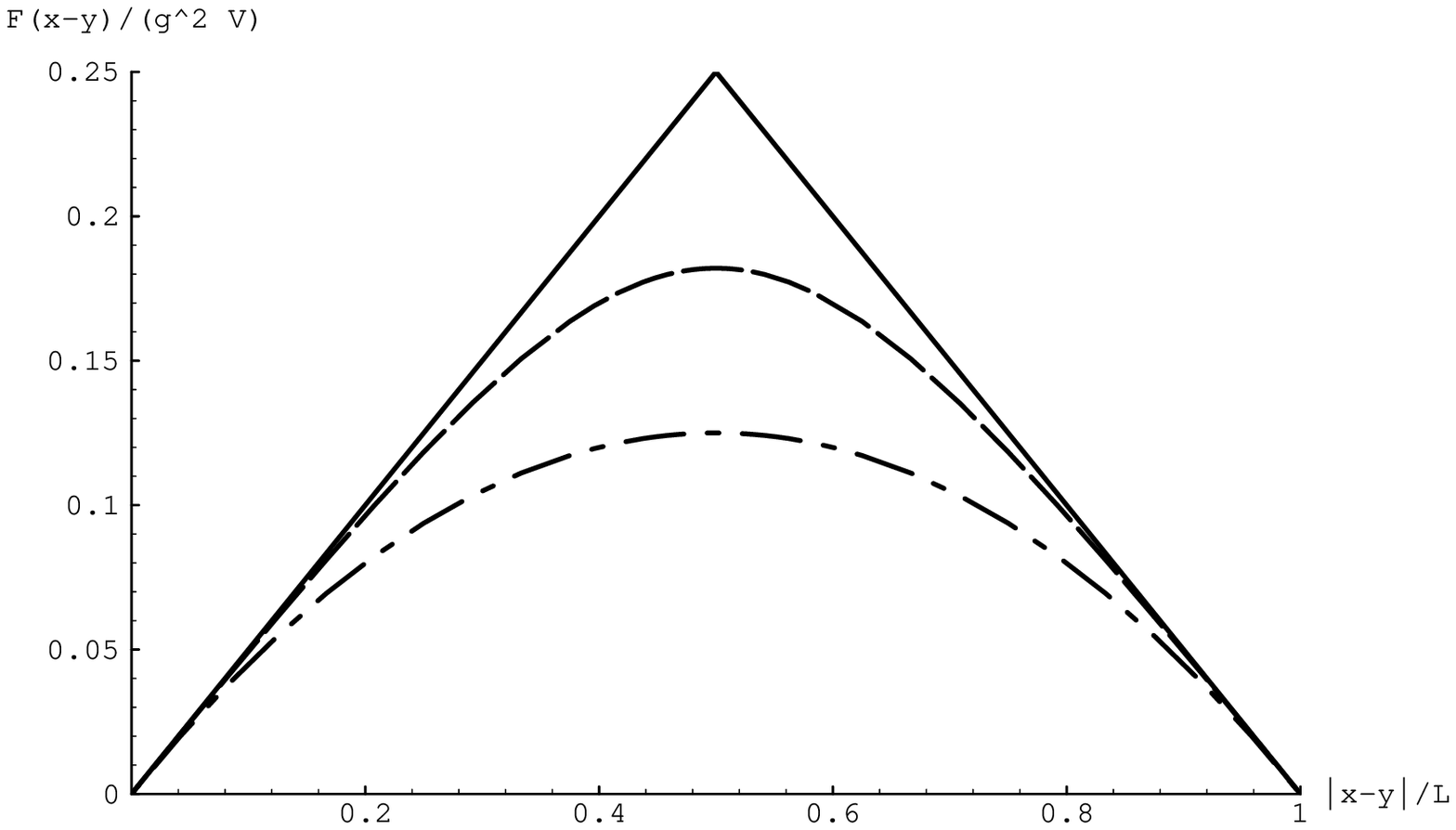}}
\end{picture}
\begin{center}{\bf Figure 1:}\\\vspace{0.2cm}  $g^2 V=\infty$:
  \rule[1mm]{10mm}{0.1mm},\ \  $g^2 V\sim
  1$: $-\!-\!- -$,\ \  $g^2 V =0$: $-\cdot-\cdot-$
\end{center}
\begin{flushleft} Interaction energy of two external
  charges for $SU(2)$, untwisted case.
\end{flushleft}

\begin{picture}(18,9)
\put(0,0){
\includegraphics{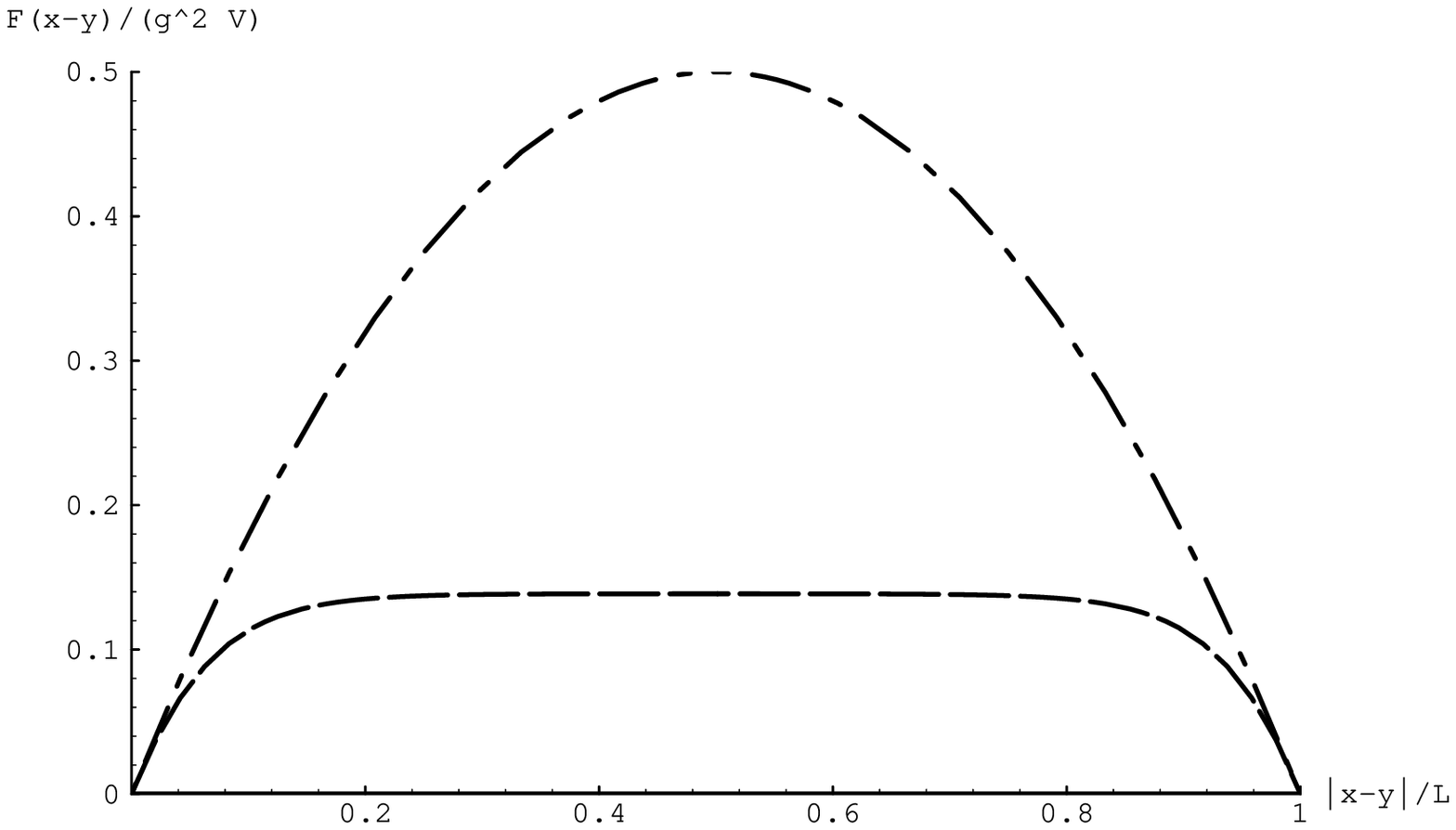}}
\end{picture}
\begin{center}{\bf  Figure 2:}\\\vspace{0.2cm} $g^2 V=\infty$:
  \rule[1mm]{10mm}{0.1mm},\ \  $g^2 V\sim
  1$: $-\!-\!- -$,\ \  $g^2 V =0$: $-\cdot-\cdot-$
\end{center} 
\begin{flushleft}  Interaction energy of two external
  charges for $SU(2)$, twisted case. $F/V=0$ for $g^2
  V=\infty$. 
\end{flushleft}

\end{document}